\documentclass[twocolumn]{aastex6}
\usepackage{graphicx}
\usepackage{amssymb,amsfonts,amsmath,amstext,amsgen,amsopn,amsxtra,indentfirst,times,longtable}

\begin{document}


\title{The Albedo Problem and Cloud Cover on Hot Jupiters}

\author{Kevin Heng\altaffilmark{1,2,3,4}}
\author{Billy Edwards\altaffilmark{5}}
\author{Nicolas B. Cowan\altaffilmark{6,7}}
\altaffiltext{1}{Faculty of Physics, Ludwig Maximilian University, Scheinerstrasse 1, D-81679, Munich, Bavaria, Germany. \\ Email: Kevin.Heng@physik.lmu.de}
\altaffiltext{2}{Munich Center for Geoastronomy, Ludwig Maximilian University, Theresienstrasse 41, D-80333, Munich, Bavaria, Germany}
\altaffiltext{3}{University College London, Department of Physics \& Astronomy, Gower St, London, WC1E 6BT, United Kingdom}
\altaffiltext{4}{Astronomy \& Astrophysics Group, Department of Physics, University of Warwick, Coventry CV4 7AL, United Kingdom}
\altaffiltext{5}{SRON, Space Research Organisation Netherlands, Niels Bohrweg 4, NL-2333 CA, Leiden, The Netherlands}
\altaffiltext{6}{Department of Earth and Planetary Sciences, McGill University, 3450 rue University, Montr\'{e}al, QC H3A OE8, Canada}
\altaffiltext{7}{Department of Physics, McGill University, 3600 rue University, Montr\'{e}al, QC H3A 2T8, Canada}

\begin{abstract}
Observations of transiting hot Jupiters have revealed a mismatch between the values of the Bond versus geometric albedos.  In the planetary science literature, the ratio of these quantities is known as the phase integral.  It has been extensively measured for the Solar System planets and shown to generally be non-unity in value.  We use existing Cassini data of Jupiter to derive bandpass-integrated geometric albedos and phase integrals in the CHEOPS, TESS and Ariel bandpasses, demonstrating that these quantities vary markedly across these different wavelength ranges.  By performing a population study of geometric albedos and phase integrals, we demonstrate that atmospheres with partial cloud cover may be identified using measurements of the phase integral if its measured uncertainty is $\sim 0.1$, which corresponds to an uncertainty of $\sim 3\%$ on the optical/visible secondary eclipse depth.  The upcoming Ariel space mission will conduct an unprecedented statistical survey of cloud cover on hot Jupiters via the simultaneous measurement of $\sim 100$ infrared phase curves and optical secondary eclipses.  Whenever available, the shape of optical phase curves of reflected light will directly constrain the phase integral, spherical albedo, degree of cloud cover and scattering asymmetry factor.
\end{abstract}


\section{Introduction}
\label{sect:intro}

In the exoplanet literature, there is a long-standing conundrum concerning the Bond and geometric albedos of hot Jupiters: the former has an average value of 0.35 compared to the latter with an average value of 0.1 \citep{sc15}.  Bond albedos quantify the reflectivity of a hot Jovian atmosphere over all viewing angles and wavelengths, and are derived from infrared phase curves, which measure the thermal emission of a hot Jupiter, using zero-dimensional ``box models" \citep{ca11}.  Geometric albedos are wavelength-dependent quantities that record the reflectivity of a hot Jovian atmosphere at superior conjunction (zero phase angle), and are derived from the optical/visible secondary eclipse depths of hot Jupiters \citep{hd13}.

In the planetary science (Solar System) literature, the ratio of the Bond albedo to the geometric albedo is known as the ``phase integral", which is usually denoted as $q$ \citep{russell1916,hapke81,pearl90,pc91}.  It has been measured for Jupiter, Saturn, Neptune and Uranus; typically, we have $q \ne 1$ \citep{pc91,li18}.  The phase integral encodes information about the properties of the reflector \citep{russell1916} such as particle size \citep{heng21,morris24}.  For example, Rayleigh scattering produces $q \approx 1.30$ \citep{heng21}.  As another example, a Lambertian sphere has $q=1.5$ (cf. pg. 38 of \citealt{seager10}).  In the current Letter, our first objective is to use previously measured Cassini data of Jupiter \citep{li18} to demonstrate that the bandpass-integrated geometric albedos and phase integrals differ between various observatories (CHEOPS, TESS, etc).

\begin{figure*}[!ht]
\begin{center}
\includegraphics[width=\columnwidth]{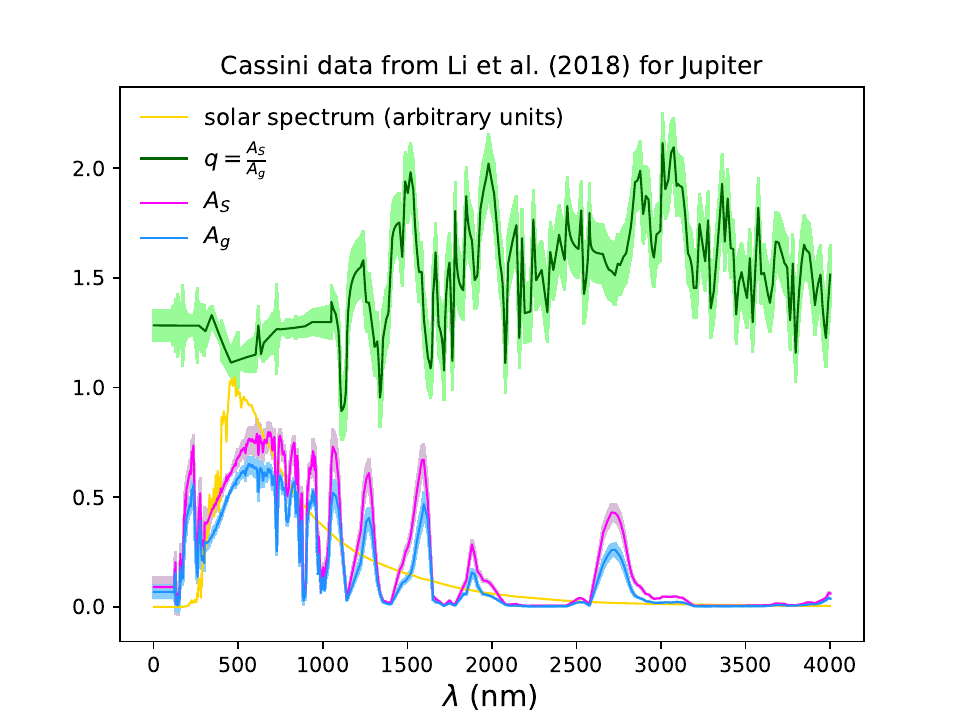}
\includegraphics[width=\columnwidth]{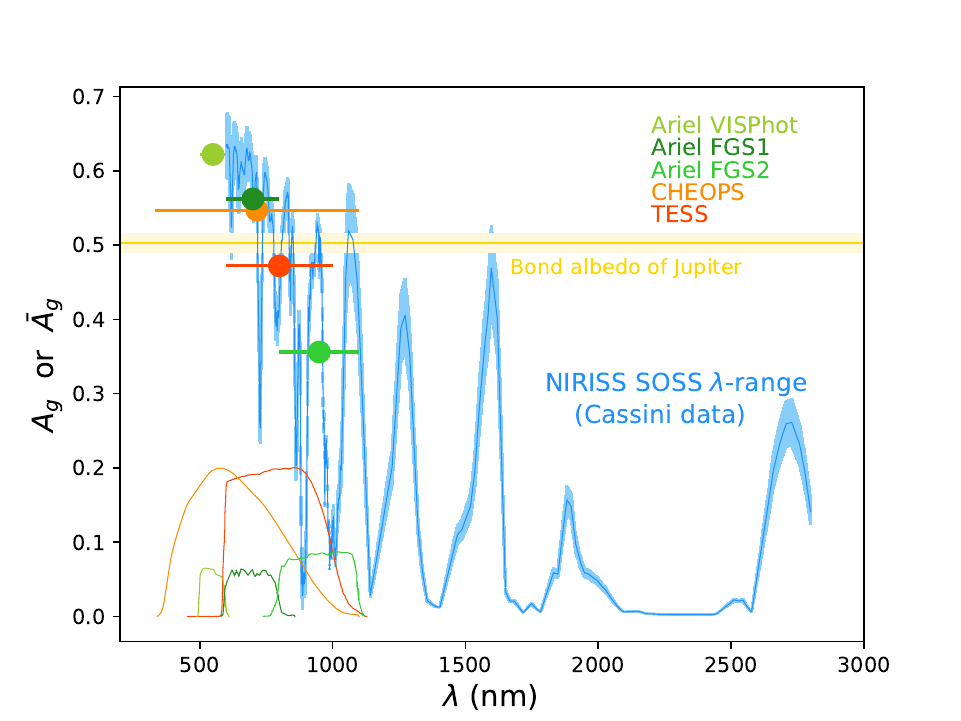}
\end{center}
\vspace{-0.1in}
\caption{Left panel: Previously measured Cassini phase integral, spherical albedo and geometric albedo of Jupiter as functions of wavelength (from \citealt{li18}).  Right panel: Bandpass-integrated geometric albedos computed using the respective filter response functions and the geometric albedo spectrum measured by the Cassini spacecraft.  For comparison, we plot the Cassini geometric albedo spectrum across the JWST NIRISS SOSS wavelength range and the spectral response functions for Ariel, CHEOPS and TESS.}
\label{fig:jupiter}
\end{figure*}

Our second and main objective is to demonstrate that the phase integral provides largely model-independent, empirical constraints on cloud cover \citep{wong21}.  Inhomogeneous cloud cover was first detected on the hot Jupiter Kepler-7b via the westward peak offset of its reflected light phase curve \citep{demory13,hu15}.  Its phase integral was the first ever measured for an exoplanet: $1.77 \pm 0.07$, integrated over the Kepler bandpass \citep{heng21}.  Inhomogeneous cloud cover has since been detected for the hot Jupiters Kepler-41b and KOI-13b \citep{morris24}, as well as the ultra-hot exoplanets WASP-121b \citep{splinter25} and LTT 9779b \citep{coulombe25}.  Therefore, the ``albedo problem" \citep{crossfield15,sc15,splinter25} amounts to $q \ne 1$ and is actually a diagnostic for one of the most important properties of clouds/hazes on hot Jupiters.

In Section \ref{sect:methods}, we clarify formalism concerning the phase integrals and geometric albedos and describe our approach for computing them.  In Section \ref{sect:results}, we present bandpass-integrated geometric albedos and phase integrals from Jupiter, as well as a population study of these quantities.  In Section \ref{sect:discuss}, we discuss the implications of our findings including the requirement on the measured uncertainty, associated with the phase integral, in order to conduct a statistical survey of cloud cover on hot Jupiters.

\section{Methodology}
\label{sect:methods}

\begin{figure*}[!ht]
\begin{center}
\includegraphics[width=\columnwidth]{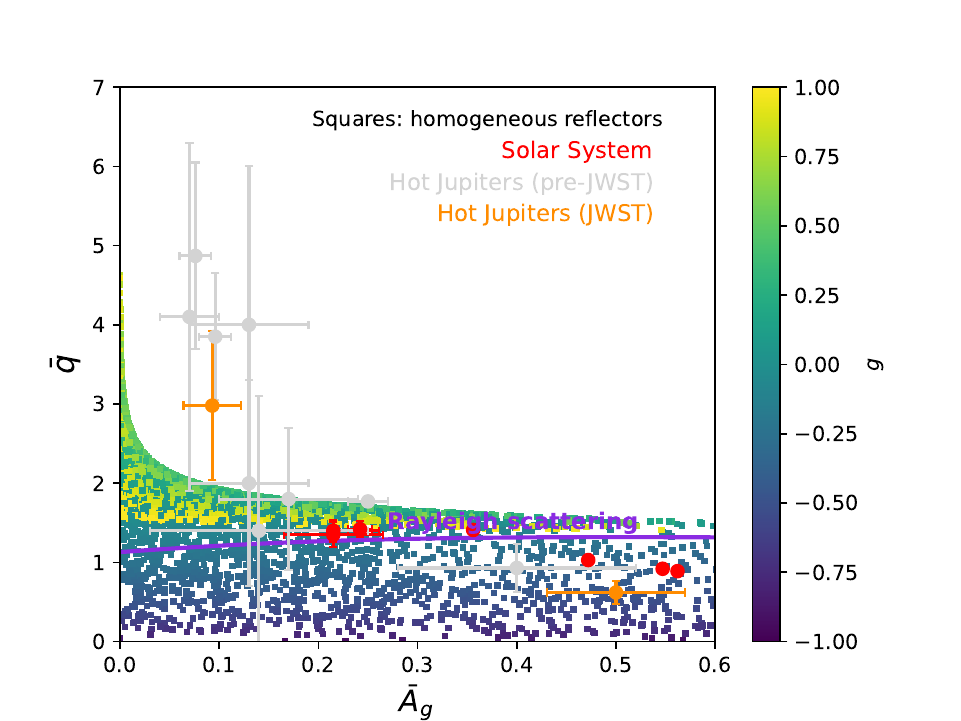}
\includegraphics[width=\columnwidth]{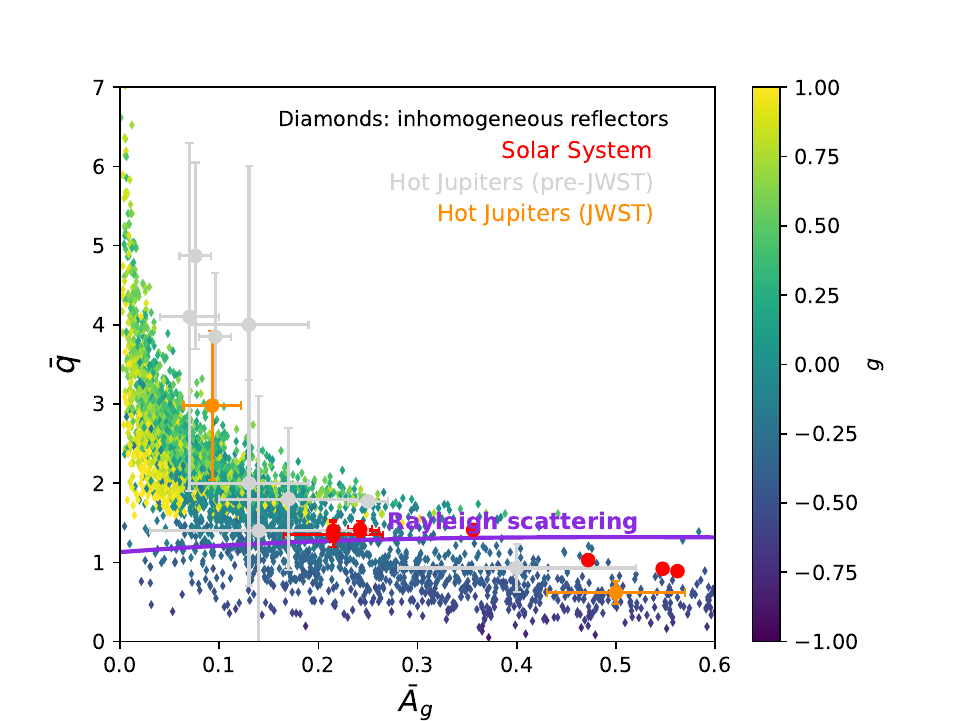}
\end{center}
\vspace{-0.1in}
\caption{Bandpass-integrated phase integrals versus geometric albedos for both real targets and synthetic populations.  The real targets include the Solar System gas/ice giants and exoplanets both in the pre-JWST era and using JWST (curated data listed in Table 1).  As the data are only for display purposes, we plot symmetric error bars using the larger of the uncertainties when they are asymmetric.  For the synthetic population (5000 random draws), we have computed $q$ using the mathematical solutions of \cite{heng21} as described in the text.  The left panel shows a synthetic population of homogeneous reflectors (2 parameters), while the right panel shows one of inhomogeneous reflectors (5 parameters).  In both cases, a Henyey-Greenstein scattering phase function is assumed and only the randomly generated population of scattering asymmetry factors ($g$) is displayed in color.}
\label{fig:pop}
\end{figure*}

\subsection{Clarifying formalism}

The geometric albedo $A_g(\lambda)$ is a wavelength-dependent quantity, i.e., one formally measures the geometric albedo \textit{spectrum}.  In practice, one typically measures the \textit{bandpass-integrated} geometric albedo,
\begin{equation}
\bar{A}_g = \frac{\int ~A_g ~I_\star ~S_\lambda ~d\lambda}{\int ~I_\star S_\lambda ~d\lambda},
\label{eq:abar}
\end{equation}
where $I_\star(\lambda)$ is the spectral energy distribution (intensity) of the star and $\lambda$ denotes the wavelength.  For photon counters (e.g., CHEOPS, optical channels of Ariel), the weighting factor is $S_\lambda(\lambda) = \lambda f_\lambda(\lambda)$ where $f_\lambda(\lambda)$ is the filter response function of the detector.  The factor of $\lambda$ converts $I_\star(\lambda)$ from flux into photon count units.  For energy counters, we have $S_\lambda(\lambda) = f_\lambda(\lambda)$.

Bandpass-integrated geometric albedos have long been measured for the gas and ice giants of the Solar System, as listed in Table 1.  For exoplanets, bandpass-integrated geometric albedos are standard practice as measured using the MOST \citep{rowe06,rowe08}, CoRoT \citep{snellen09}, Kepler \citep{hd13,demory14,esteves15}, TESS \citep{wong21} and CHEOPS (e.g., \citealt{brandeker22,krenn23}) space telescopes.  Most of these measurements are of hot Jupiters.

The phase integral $q(\lambda)$ is strictly a wavelength-dependent quantity as well.  It is the ratio of the spherical to geometric albedos \citep{russell1916,sobolev},
\begin{equation}
q = \frac{A_S}{A_g},
\label{eq:q}
\end{equation}
where the spherical albedo $A_S(\lambda)$ is wavelength-dependent and considers all viewing angles \citep{seager10}.  It is not the spherical albedo, but the \textit{shape} of the reflected light phase curve, known as the integral phase function $\Psi(\alpha,\omega,g)$ \citep{russell1916,hapke81}, that is directly measured from the phase curve \citep{heng21}.  Denoting the orbital phase angle by $\alpha$, the phase integral is \citep{russell1916,hapke81,heng21}
\begin{equation}
q  = \int^{\pi}_{-\pi} \Psi ~\sin{\vert \alpha \vert} ~d\alpha.
\end{equation}
Once $A_g$ and $q$ are constrained by the data, one may derive $A_S$ using equation (\ref{eq:q}).  The Bond albedo is the stellar spectrum-weighted average spherical albedo \citep{seager10}:  
\begin{equation}
A_B = \frac{\int ~A_S ~I_\star ~d\lambda}{\int ~I_\star ~d\lambda}.
\label{eq:bond}
\end{equation}
In other words, $A_g$ and $\Psi$ are the magnitude and shape of a reflected light phase curve, respectively \citep{seager10}.

The bandpass-integrated phase integral is formally defined as
\begin{equation}
\bar{q} = \frac{\int ~q ~I_\star ~S_\lambda ~d\lambda}{\int ~I_\star ~S_\lambda ~d\lambda}.
\label{eq:qbar_real}
\end{equation}
In practice, the quantity in equation (\ref{eq:qbar_real}) is rarely measured.  It has been measured for Jupiter \citep{li18} and the hot Jupiter Kepler-7b \citep{heng21} using data from the Cassini and Kepler missions, respectively.  

Rather, the bandpass-integrated phase integral is commonly approximated as (e.g., \citealt{pearl90,pc91,wong21})
\begin{equation}
\bar{q} \approx \frac{A_{\rm B}}{\bar{A}_g}.
\label{eq:qbar}
\end{equation}
It is apparent that equation (\ref{eq:qbar}) does not follow mathematically from integrating equation (\ref{eq:q}) over wavelength and using equations (\ref{eq:abar}) and (\ref{eq:bond}).  \textit{Integrating over a ratio is not equal to taking the ratio of two integrals.}  In other words, the quantity in equation (\ref{eq:qbar}) is not the true bandpass-integrated geometric albedo inferred from the shape of a reflected light phase curve.  Values of $\bar{q}$ for Solar System planets and hot Jupiters are again tabulated in Table 1.  

\subsection{Calculating the geometric albedo and phase integral}
\label{subsect:theory}

The geometric albedo ($A_g$) and phase integral ($q$)  may be calculated using the ab initio mathematical solutions of \cite{heng21}.  These solutions build on the seminal work of \cite{chandra60}, who derived an exact solution for the emergent/reflected intensity from a semi-infinite\footnote{Infinite across optical depth and not distance, meaning the atmosphere transitions from being transparent to being opaque.} atmosphere with constant scattering properties at a specific wavelength, which imply a spatially uniform population of particles but not necessarily with the same size \citep{hl21}.  \cite{hapke81} derived an accurate, approximate solution for the Chandrasekhar $H$-functions using the two-stream approximation, which significantly speeds up the computation of the intensity.  \cite{heng21} used these developments to derive analytical expressions for $A_g$ and $q$, although the latter requires the numerical evaluation of an integral.  These mathematical solutions make one approximation: multiple scattering occurs isotropically \citep{hapke81}, but single scattering may be described by a scattering phase function of arbitrary functional form \citep{heng21}.

The solutions of \cite{heng21} come in two flavors.  The first is for a homogeneous reflector, which is an atmosphere that has the same reflectivity globally.  In such a model, the parameters are the single scattering albedo $\omega(\lambda)$ and the scattering asymmetry factor $g(\lambda)$.  The second is for an inhomogeneous reflector (see schematic in Figure 1 of \citealt{hu15}), which is an atmosphere that is homogeneous in latitude but has dark versus bright regions across longitude.  Denoting the local longitude by $x$, where $x=0^\circ$ corresponds to the substellar point, the dark (poorly reflective) region of the atmosphere is bounded between the longitudes $x_1$ and $x_2$, where $-90^\circ < x_1 < x_2 < 90^\circ$, while the rest of the atmosphere is bright (highly reflective).  Such a setup mimics a ``patchy cloud" configuration \citep{hu15,lp16,oreshenko16}, where the bright region corresponds to cooler parts of the atmosphere that have temperatures below the condensation temperature of the cloud.  The dark and bright regions have single-scattering albedos of $\omega_0$ and $\omega = \omega_0 + \omega^\prime$, respectively.  

The Henyey-Greenstein scattering phase function or reflection law is assumed \citep{hg41}, but in principle the practitioner is free to use any scattering phase function as the mathematical derivation of \cite{heng21} is general.  Degeneracies (between $g$, $\omega$, $x_1$ and $x_2$) associated with fitting the model of the inhomogeneous reflector to a reflected light phase curve have previously been discussed in \cite{heng21} and \cite{morris24} for several case studies.

\section{Results}
\label{sect:results}

\subsection{Jupiter as an exoplanet}
 
Exquisite data of Jupiter have previously been obtained using the Cassini spacecraft \citep{li18}.  Figure \ref{fig:jupiter} shows $q(\lambda)$, $A_S(\lambda)$ and $A_g(\lambda)$ of Jupiter.  These data have been used to study Jupiter as an exoplanet \citep{hl21}.  To demonstrate that the bandpass-integrated geometric albedos differ between the different wavelength ranges probed by various observatories, we use the Cassini-measured $A_g(\lambda)$ to derive $\bar{A}_g$ and $\bar{q}$ using the respective filter response functions (for CHEOPS, TESS and Ariel) and equations (\ref{eq:abar}) and (\ref{eq:qbar}), respectively.  The spectral energy distribution of the Sun $I_\star = I_\odot(\lambda)$, as compiled by \citep{li18}, is used.  To compute $\bar{q}$, we used the measured Bond albedo of $A_{\rm B} = 0.503 \pm 0.012$ \citep{li18}.  Figure \ref{fig:jupiter} shows that the bandpass-integrated geometric albedos and phase integrals have a range of values of $\bar{A}_g \approx 0.36$--$0.62$ and $\bar{q} \approx 0.8$--$1.4$, respectively (Table 1).

Previously, the TESS geometric albedo of Jupiter was calculated to be $\bar{A}_g = 0.489 \pm 0.003$ \citep{wong21}.  The website of the Spanish Virtual Observatory (SVO) Filter Profile Service\footnote{http://svo2.cab.inta-csic.es/theory/fps/} \citep{rodrigo24} explicitly lists the TESS filter response function as an energy counter, which motivated the use of $S_\lambda = f_\lambda$ when using equation (\ref{eq:abar}) for computation \citep{wong21}.  We have since verified that the TESS filter response function is for a photon counter \citep{cubillos25}, which instead requires $S_\lambda = \lambda f_\lambda$.  This correction slightly revises the TESS geometric albedo of Jupiter to be $\bar{A}_g = 0.472 \pm 0.003$ instead.
 
\subsection{Population study of phase integrals and cloud cover}

Figure \ref{fig:pop} displays a collection of bandpass-integrated phase integrals and geometric albedos of Solar System planets (Jupiter, Saturn, Neptune and Uranus) and exoplanets as curated in Table 1.  Overplotted is a curve associated with Rayleigh scattering by a homogeneous atmosphere, calculated using the analytical formulae of \cite{heng21}.  The bandpass-integrated geometric albedos of Jupiter (three entries shown in Figure \ref{fig:pop}), WASP-121b and LTT 9779b are inconsistent with Rayleigh scattering by a homogeneous reflector.  For Jupiter, this is because the cloud particles or aerosols are large and polydisperse \citep{hl21}.  For WASP-121b \citep{splinter25} and LTT 9779b \citep{coulombe25}, this is because inhomogeneous cloud cover has been detected from the westward peak offsets of their optical phase curves.  The cloud particles in the Jovian atmosphere are unlikely to be representative of those in hot Jovian atmospheres, because the chemical and temperature conditions, as well as the dynamical structure, of Jupiter differ from those of hot Jupiters.

\begin{figure}[!ht]
\begin{center}
\vspace{-0.1in} 
\includegraphics[width=\columnwidth]{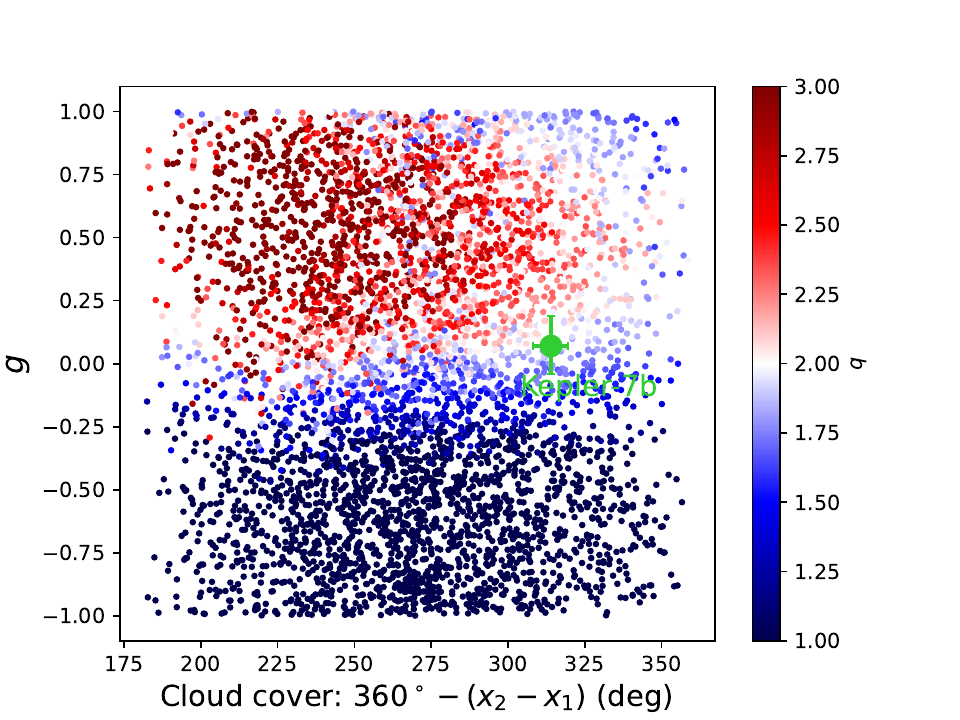}
\end{center}
\vspace{-0.1in}
\caption{Population study of the phase integral using the same synthetic population of inhomogeneous reflectors generated in Figure \ref{fig:pop}, but plotted as functions of the scattering asymmetry factor and degree of cloud cover.  For display purposes, we have restricted the color range for $q=1$ to $3$.  Shown are the measured degree of cloud cover and scattering asymmetry factor for Kepler-7b (see text for details).}
\label{fig:cover}
\end{figure}

To generate a synthetic population of homogeneous and inhomogeneous reflectors, we randomly sample the parameters needed for calculation.  For homogeneous reflectors, we uniformly sample the single-scattering albedo from $\omega=0$ to $1$.  For inhomogeneous reflectors, we uniformly sample $\omega_0=0$ to $0.4$ (as motivated by the findings of \citealt{morris24}), $x_1=-90^\circ$ to $0^\circ$ and $x_1=0$ to $90^\circ$.  We then have $\omega^\prime = 1-\omega_0$ such that the single-scattering albedo of the bright region is unity.  For both homogeneous and inhomogeneous reflectors, we uniformly sample the scattering asymmetry factor \citep{hg41} $g=-1$ to $1$.

Figure \ref{fig:pop} shows that homogeneous reflectors rarely produce $q>2$.  To produce $q>1$, non-zero values of $g$ corresponding to particles larger than the wavelength of light scattered are needed.  Inhomogeneous reflectors with $g \ne 0$ easily produce $q>2$.  Figure \ref{fig:cover} visualises the same synthetic population of inhomogeneous reflectors in a different way.  For $g \ge 0$, it is apparent that $q$ varies from about $2$ to $3$ as the degree of cloud cover decreases.  

Figure \ref{fig:pop} also shows the measured cloud cover ($360^\circ - x_2+x_1 = 314^{+5.5}_{-6.0}$ deg) and scattering asymmetry factor ($g=0.07^{+0.12}_{-0.11}$) of Kepler-7b as derived by \cite{heng21} using the model (and its caveats) summarised in Section \ref{subsect:theory}.  It is worth emphasizing that the phase integral of Kepler-7b ($\bar{q} = 1.77 \pm 0.07$) was directly inferred from the \textit{shape} of the reflected light phase curve.  The position of Kepler-7b in Figure \ref{fig:cover} is consistent with the estimated value of its phase integral as calculated using the ensemble approach, which suggests that the value of the phase integral alone may provide constraints on the degree of cloud cover and the particle size (via the value of the scattering asymmetry factor).

Figure \ref{fig:cover} shows that $q$ varies from about 1 to 3.  In some regions of the parameter space of $g$ versus degree of cloud cover, $q$ varies rather steeply (by $\sim 0.1$).  This suggests that a precision on the bandpass-integrated phase integral of $\delta \bar{q} \sim 0.1$ is needed to constrain the corresponding $g$ value and degree of cloud cover, if only an optical secondary eclipse measurement is available.  If an optical phase curve of reflected light is available, then $g$ and the degree of cloud cover may be directly inferred from its shape, as has been accomplished for Kepler-7b \citep{heng21}.

\section{Discussion}
\label{sect:discuss}

\subsection{Caveats and limitations}

It is worth discussing the limitations of this ensemble approach.  Even for Kepler-7b, one could not definitively rule out the presence of homogeneous cloud cover from the values of its bandpass-integrated phase integral ($\bar{q} = 1.77 \pm 0.07$) and geometric albedo ($\bar{A}_g = 0.25^{+0.01}_{-0.02}$) alone without detecting the westward peak offset of its optical phase curve.  

Although the hot Jupiters Kepler-41b and KOI-13b have optical phase curves measured by the Kepler space telescope, they do not attain the precision of the optical phase curve of Kepler-7b, which has the consequence that the inferred scattering asymmetry factors are prior-dominated for these two objects, despite the detection of inhomogeneous cloud cover \citep{morris24}.  

In general, it is challenging to infer the functional form of the scattering phase function or reflection law from an optical phase curve, e.g., rule out a Lambertian sphere, unless its shape is measured precisely.  From an optical secondary eclipse alone, this is impossible.  It has been demonstrated that there is essentially no chance of detecting the opposition surge effect, which is observed in the planets and moons of the Solar System, from optical phase curves in the foreseeable future \citep{jones25}.  However, if optical phase curves were precisely measured at several different wavelengths, one could obtain constraints on the size distribution of cloud particles as has been demonstrated for Jupiter using Cassini data \citep{hl21}.

\subsection{The Ariel mission}

The Ariel space mission of ESA is poised to study the atmospheres of 500 to 1000 exoplanets, simultaneously covering optical and near-infrared wavelengths from 0.5 to 7.8 $\mu$m \citep{tinetti18}.  It is expected to deliver high-quality infrared eclipses for about 300 exoplanets \citep{edwards_tinetti} and a portion of these observations will provide meaningful constraints on reflected light (e.g., \citealt{zellem_case}). Work is ongoing to define the size and scope of an Ariel phase curve survey \citep{charnay_pc}, but current efforts suggest that it should be possible to measure 50 to 100 phase curves that deliver high signal-to-noise ratio (SNR) optical and infrared data, thereby providing a statistical survey of the phase integral and therefore cloud cover on hot Jupiters. 

Such a survey will offer important empirical clues on how clouds form in hot Jovian atmospheres. While optical phase curves of reflected light are capable of directly constraining the degree of cloud cover and single-scattering albedos (of the bright/cloudy and dark/cloudfree regions of an atmosphere), they are generally more demanding to measure than infrared phase curves of thermal emission. Therefore, it is possible that Ariel could measure $\sim 100$ additional phase curves that have a high SNR in the infrared, thereby constraining the Bond albedo, but a much more modest SNR in the optical.

The results in Figure \ref{fig:cover} demonstrate that one can perform a statistical survey of cloud cover if the phase integral is measured precisely enough.  Making the conservative approximation that the uncertainties on $A_{\rm B}$ and $\bar{A}_g$ are uncorrelated, one may write
\begin{equation}
\delta \bar{q} = \bar{q} \sqrt{ \left( \frac{\delta A_{\rm B}}{A_{\rm B}} \right)^2 + \left( \frac{\delta \bar{A}_g}{\bar{A}_g} \right)^2 }.
\label{eq:dq}
\end{equation}
The uncertainty on the Bond albedo may be estimated by first writing an expression for $A_{\rm B}$ in terms of the dayside temperature using zero-dimensional ``box models" \citep{ca11,morris22},
\begin{equation}
A_{\rm B} = 1 - \frac{1}{f} \left( \frac{T_{\rm day}}{T_{\rm irr}} \right)^4,
\end{equation}
where $f$ is the redistribution factor, $T_{\rm day}$ is the dayside temperature and $T_{\rm irr}$ is the irradiation temperature.  The preceding expression allows us to write
\begin{equation}
\frac{\delta A_{\rm B}}{A_{\rm B}} \approx 4 \left( \frac{\delta T_{\rm day}}{T_{\rm day}} \right),
\end{equation}
if the uncertainties on $f$ and the stellar/orbital properties are ignored.  If the measured uncertainty on the dayside temperature is $\delta T_{\rm day}/T_{\rm day} \sim 1\%$ (i.e., an uncertainty of $\sim 10$ K for $T_{\rm day} \sim 1000$ K), then the uncertainty on the Bond albedo is $\delta A_{\rm B}/A_{\rm B} \sim 4\%$. Simultaneously fitting for the Bond albedo and day-to-night heat transport leads to somewhat greater uncertainties \citep{splinter25}.  Using equation (\ref{eq:dq}), the required uncertainty on the geometric albedo (and hence the optical/visible secondary eclipse depth) is $\delta \bar{A}_g/\bar{A}_g \sim 3\%$ if $\delta \bar{q} \sim 0.1$ and $\bar{q} \sim 2$.  In practice, the need to separate out the reflected light versus thermal emission components \citep{hd13} will further increase the measured uncertainties on $A_{\rm B}$ and $\bar{A}_g$.

The key message of the current Letter is \textit{not} to ignore the rich information encoded in optical phase curves of reflected light.  It has been demonstrated that the \textit{shape} of reflected light phase curves allows the bandpass-integrated phase integral, spherical albedo and scattering asymmetry factor to be directly constrained \citep{heng21,morris24}.  However, measuring the shape of reflected light phase curves to sufficient precision requires an expensive investment of telescope time.  If faced with a dearth of \textit{precise} optical phase curves, the current study demonstrates that an ensemble of optical secondary eclipses, when combined with Bond albedos inferred from thermal phase curves, may provide population-level constraints on exoplanet cloud cover.

\begin{table*}
\label{tab:bandpass}
\begin{center}
\caption{Bandpass-integrated geometric albedos and phase integrals}
\begin{tabular}{lcccc}
\hline
\hline
Object & Bandpass & $\bar{A}_g$ & $\bar{q}$ & Reference \\
\hline
Jupiter & Ariel VISPhot (0.5--0.6 $\mu$m) & $0.622 \pm 0.008$ & $0.81 \pm 0.03$ & This work \\
Jupiter & Ariel FGS1 (0.6--0.8 $\mu$m) & $0.562 \pm 0.003$ & $0.89 \pm 0.03$ & This work \\
Jupiter & Ariel FGS2 (0.8--1.1 $\mu$m) & $0.356 \pm 0.004$ & $1.41 \pm 0.03$ & This work \\
Jupiter & CHEOPS (0.33--1.1 $\mu$m) & $0.547 \pm 0.003$ & $0.92 \pm 0.03$ & This work \\
Jupiter & TESS (0.6--1.0 $\mu$m) & $0.472 \pm 0.003$ & $1.07 \pm 0.03$ & This work \\
Saturn & Voyager 2 (0.3--1.9 $\mu$m) & $0.242 \pm 0.012$ & $1.42 \pm 0.10$ & \cite{hanel83,pc91} \\
Uranus & Voyager 2 (0.3--1.9 $\mu$m) & $0.215 \pm 0.046$ & $1.40 \pm 0.14$ & \cite{pearl90,pc91} \\
Neptune & Voyager 2 (0.3--1.9 $\mu$m) & $0.215 \pm 0.050$ & $1.35 \pm 0.16$ & \cite{pc91} \\
Kepler-7b & $^\ddagger$Kepler (0.42--0.9 $\mu$m) & $0.25^{+0.01}_{-0.02}$ & $1.77 \pm 0.07$ & \cite{heng21} \\
CoRoT-2b & $^\dagger$CoRoT (0.35--1.0 $\mu$m) & $0.07 \pm 0.03$ & $4.1 \pm 2.2$ & \cite{wong21} \\
Qatar-1b & $^\dagger$TESS (0.6--1.0 $\mu$m) & $0.14 \pm 0.11$ & $1.4 \pm 1.7$ & \cite{wong21} \\
WASP-12b & $^\dagger$TESS (0.6--1.0 $\mu$m) & $0.13 \pm 0.06$ & $2.0 \pm 1.3$ & \cite{wong21} \\
WASP-19b & $^\dagger$TESS (0.6--1.0 $\mu$m) & $0.17 \pm 0.07$ & $1.8 \pm 0.9$ & \cite{wong21} \\
WASP-43b & $^\dagger$TESS (0.6--1.0 $\mu$m) & $0.13 \pm 0.06$ & $4.0 \pm 2.0$ & \cite{wong21} \\
HD 189733b & $^\dagger$CHEOPS (0.33--1.1 $\mu$m) & $0.076 \pm 0.016$ & $4.87^{+1.13}_{-1.18}$  & \cite{sc15,krenn23} \\
HD 189733b & $^\dagger$HST (0.29--0.45 $\mu$m) & $0.40 \pm 0.12$ & $0.93^{+0.29}_{-0.30}$ & \cite{evans13,sc15} \\
HD 209458b & $^\dagger$CHEOPS (0.33--1.1 $\mu$m) & $0.096 \pm 0.016$ & $3.85^{+0.75}_{-0.80}$  & \cite{sc15,brandeker22} \\
LTT 9779b$^\spadesuit$ & NIRISS SOSS (0.6--2.85 $\mu$m) & $0.50 \pm 0.07$ & $0.62 \pm 0.15$ & \cite{coulombe25} \\
WASP-121b & NIRISS SOSS (0.6--2.85 $\mu$m) & $0.093^{+0.029}_{-0.027}$ & $2.98^{+0.94}_{-0.88}$ & \cite{splinter25} \\
\hline
\hline
\end{tabular}\\
$\dagger$: Optical/visible eclipses from stated observatory, Bond albedo from Spitzer phase curves. \\
$\ddagger$: Only true exoplanet phase integral derived entirely from reflected light.\\
$\spadesuit$: Hot Neptune with $R \approx 4.7 R_\oplus$.
\vspace{0.1in}
\end{center}
\end{table*}

\vspace{0.2in}
{\scriptsize This project stemmed from coffee break discussions at the 2025 Ariel Summer School (Fréjus, France), where all three co-authors served as lecturers, and the initial manuscript was written in a 48-hour push at the 2025 Ariel Consortium Meeting (Madrid, Spain).  KH developed the theoretical formalism, derived the equations, performed the numerical calculations, made the figures, led the writing of the manuscript and fed off the energy of BE and NBC. BE added albedos to Table 1, provided the Ariel filter response functions and perspective from simulations of Ariel observations, and contributed to writing the manuscript. NBC curated some of the albedos in Table 1, performed a literature review of albedos measured by JWST, provided general perspective on the measured uncertainties and contributed to writing the manuscript. KH acknowledges partial financial support from the European Research Council (ERC) Geoastronomy Synergy Grant (grant number 101166936; PIs: Gaillard, Heng and Mojzsis).  NBC acknowledges support from a Canada Research Chair and NSERC Discovery Grant. He is also grateful to the Trottier Space Institute and l’Institut de recherche sur les exoplanètes for their financial support and dynamic intellectual environment.}

\label{lastpage}

\end{document}